\documentclass[10pt]{article} 
\usepackage{epsf,epsfig,graphicx}
\usepackage{times}

\newcommand*{\figref}[1]{Fig.\ref{#1}}

\begin{document}

\twocolumn[{\LARGE \textbf{Defect formation of lytic peptides in lipid
membranes and their influence on the thermodynamic properties of the
pore environment.\\*[0.1cm]}}\\

{\large Vitaliy Oliynyk$^{1,2}$, Udo Kaatze$^1$ and Thomas 
Heimburg,$^{2,\ast}$\\*[0.1cm] 
{\footnotesize $^1$Complex Fluids Group, Drittes Physikalisches Institut,
Georg-August Universit\"at, Friedrich-Hund-Platz 1, D-37077
G\"ottingen, Germany\\
$^2$The Niels Bohr Institute,
University of Copenhagen, Blegdamsvej 17, 2100 Copenhagen \O,
Denmark\\*[0.3cm]}

{\normalsize We present an experimental study of the pore formation
processes of small amphipathic peptides in model phosphocholine lipid
membranes.  We used atomic force microscopy to characterize the
spatial organization and structure of alamethicin- and melittin-
induced defects in lipid bilayer membranes and the influence of the
peptide on local membrane properties.  Alamethicin induced holes in
gel DPPC membranes were directly visualized at different peptide
concentrations.  We found that the thermodynamic state of lipids in
gel membranes can be influenced by the presence of alamethicin such
that nanoscopic domains of fluid lipids form close to the peptide
pores, and that the elastic constants of the membrane are altered in
their vicinity.  Melittin-induced holes were visualized in DPPC and
DLPC membranes at room temperature in order to study the influence of
the membrane state on the peptide induced hole formation.  Also
differential scanning calorimetry was used to investigate the effect
of alamethicin on the lipid membrane phase behavior.\\*[0.5cm]

}}
]

\section{Introduction}
The biological activities of membrane active peptides are determined
to a large part by their interactions with the phospholipid bilayer
comprising the plasma membrane and the mutual structural effects
induced within the peptide and lipid molecules. Many native and
synthetic peptides are known to, under certain conditions, form
spontaneously transmembrane defects, such as pores, in lipid
bilayers. Defect formation is promoted by a combination of
electrostatic interactions of the
peptide residues with the polar heads of anionic lipid molecules and
hydrophobic interactions with the lipid acyl
chains~\cite{Spach1989,Shai1999,Kourie2000}. It is commonly believed
that the defect formation is the mode of action of antimicrobial
peptides. These peptides are active, forming pores, in some cell
membranes but not in others, such that they can function as
host-defense agents,
killing microbes. On certain conditions, however, they are just
associated to the membrane surface where they are inactive. Studies
of defect formation and of the defect structures in lipid bilayers
may be a key
step toward our understanding of how the activities of antimicrobial
peptides are regulated in the biological world.

Amphipathic, $\alpha$-helical peptides are abundant in nature, serving
as membrane permeating agents in the host defense system of many
organisms.  Antibiotic peptides, such as alamethicin, isolated from
the \textit{Trichoderma viride fungus}, and the bee venom peptide
melittin are among the most intensively studied
peptides~\cite{Cafiso1994,Dempsey1990}.  As the amino-acid sequence
and the helical structure in water as well as membrane environments of
these peptides is well known, they can serve as convenient models for
studies of interactions between membrane located proteins and lipids.
The crystal structures of alamethicin and melittin have been solved
more than twenty years ago by X-ray
crystallography~\cite{Fox1982,Terwilliger1982,Gennis1989}.  At low
peptide-to-lipid molar ratios alamethicin preferentially adsorbs to
the membrane surface where it is arranged parallel to the lipid
headgroups and associated to the bilayer surface.  With increasing
peptide concentration, alamethicin switches to an active state.  It is
then inserted into the lipid membrane, forming transmembrane
pores~\cite{He1996a,He1996}.  Above a certain critical concentration
nearly all peptide molecules are participating in the pore formation
process~\cite{Zuckermann2001,Chen2002}.  The structure of alamethicin
channels is generally considered in terms of the "barrel-stave"
model~\cite{Hall1984,Duclohier1992,Laver1994,Cantor2002,Duclohier2001},
in which multiple peptide molecules form a helix bundle surrounding a
central pore.  This model is capable to explain the occurrence of
channel activity in discrete, multilevel conductance
steps~\cite{Eisenberg1973,Hanke1980,Keller1993,Taylor1991} which is
caused by a varying number of pore-forming peptides.  Nevertheless, in
spite of the fact, that a large body of experimental data was
generated, the microscopic structure of alamethicin pores and the
organization of peptide pores in lipid membranes is not completely
understood presently.

Proteins and peptides inserted into membranes may influence the chain
melting transition of lipid membranes. It is known from calorimetric
studies that gel-to-fluid transition profiles are broadened and/or
shifted to either lower or higher temperatures by addition of
proteins~\cite{Ivanova2001,Ivanova2003,Pedersen2005}. The shape of
the heat capacity profiles contains valuable information on the modes
of interactions between peptides and
lipids~\cite{Heimburg1996,Ivanova2001}, for example, about their
spatial organization. It has been shown that the effect of integral
peptides on the phase behaviour of lipid membranes strongly depends
on the chain length of the lipids~\cite{Zhang1995}. This finding is
discussed in terms of "hydrophobic mismatch", which
implies that the interaction between integral proteins (or
amphipathic peptides) and lipids depends on the relative length
difference of their hydrophobic cores. It has been proposed that the
hydrophobic mismatching controls the peptide partitioning in lipid
membranes via lipid mediated forces~\cite{Mouritsen1993,Jensen2004}.

Atomic force
microscopy~\cite{Binnig1986,Hansma1994,Janshoff2001,Connell2006,Kruijff2006}
is extensively used in recent studies for the characterization of
lipid membrane systems with resolution on the nanoscopic scale.  This
method was successfully applied to investigate, with high spatial
resolution, the structure of pure lipid membranes as well as peptide
containing membranes under different
conditions~\cite{Nielsen2000,Tokumasu2003,Leidy2002,Rinia2001,Kaasgaard2003}.
Direct visualization of peptide aggregates in model membranes as well
as the study of their structure and their effect on lipid bilayers was
reported for a number of native~\cite{Mou1996,Ivanova2003} and
synthetic peptides~\cite{Rinia2001,Pedersen2005}.  The experiments
reported in this study focus on the characteristics of alamethicin as
well as melittin association and aggregation within lipid membranes.
We found the appearance of pores or defects in the membrane induced by
such peptides.  We loosely refer to these features as `defects'
although they may be closely related or indistinguishable from pores.
The investigation demonstrates that the structure of alamethicin- and
melittin-induced transmembrane defects in gel membranes can be
directly visualized for different peptide concentrations.  The
influence of the peptides on the phase behaviour of phosphocholine
membranes is also reported.  In particular, we show that the physical
behavior of the lipid membrane is altered in the vicinity of the
pores.

\section{Materials and Methods}
1,2- dipalmitoyl- sn- glycero- 3- phosphatidylcholine (DPPC),
1,2- dimyristoyl- sn- glycero- 3- phosphatidylcholine (DMPC) and
1,2- dilauroyl- sn- glycero- 3- phosphocholine (DLPC) were purchased
from
Avanti Polar Lipids (Alabaster, AL). Alamethicin was provided by
Sigma (St.~Louis, MO) and melittin as a powder from Alexis
Biochemicals (San~Diego, CA). All substances were used without
further purification. Ruby muscovite mica was obtained from TED
PELLA, inc. (Redding, CA). 

For the preparation of lipid-peptide multilamellar vesicle
dispersions, lipids and peptides were separately dissolved in a 1:1
mixture of dichloromethane and methanol. The dissolving of lipids and
peptides in the organic solvents, preceding the preparation of
aqueous solutions, was required for more exact weighing of the
substances in micrograms amounts and for better mixing of peptides
and lipids. Further, appropriate amounts of the solutions of the
target substances were mixed together, dried under a weak flow of
nitrogen gas, and placed under vacuum overnight to remove the
residual solvent. The dried peptide/lipid mixtures were dispersed in
Milli-Q water to a final concentration of 1--3 mM. Aqueous
multilamellar vesicle dispersions were prepared by heating the
samples above 50$^{\circ}$C , followed by vortexing.

Differential scanning calorimetry (DSC)
experiments~\cite{Plotnikov1997} were performed using large
unilamellar lipid vesicle (LUV) suspensions. LUVs were obtained with
the aid of a small volume extrusion apparatus~\cite{MacDonald1991}
provided by Avestin (Ottawa, Ca{-}nada). The multilamellar vesicles were
extruded through polycarbonate filters with 100~nm pores size,
mounted in the mini-extruder  and fitted with two 1.0 ml syringes.
Samples were 21 times passed through the filter membrane.
An odd number of passages was performed to avoid contamination of the
sample by multilamellar vesicles which might not have passed through
the filter. During the extrusion process the temperature of the
sample was kept above the melting point of the lipids that
facilitates the pushing of the lipid suspension through the filter.
Right before the filling of the calorimeter the solution of extruded
vesicles was degassed for 15 minutes in order to remove air
microbubbles. DSC experiments were performed using a VP-DSC from
MicroCal (Northhampton, MA) on samples of 5 mM lipid LUVs at a scan
rate of 2$^{\circ}$C/h. An appropriate baseline was subtracted from the
resulting thermograms.

Samples for atomic force microscopy (AFM)
experiments ~\cite{Binnig1986} were prepared utilizing direct fusion of
small unilamellar vesicles (SUVs) on mica~\cite{Shao1995}.  Lipid SUVs
were prepared in the presence of the peptides by sonication with the
aid of a Sonifier Cell Disruptor B-15 (Branson, Germany) until the
solution became completely transparent.  Transparency ensures that the
solution consists mostly of small unilamellar vesicles.  The SUVs were
immediately rewarmed to temperature above 55$^{\circ}$C, and 40--80~$\mu$l
of the vesicle suspension was added to a piece of freshly cleaved
mica.  The samples were incubated for 20~min at room temperature and
rinsed by exchanging ten times the incubation solution with 150 mM
NaCl solution afterwards (rinsing by Milli-Q water was found to be
also effective).  In doing so the supported lipid membrane was never
allowed to dry.  The mica-supported lipid bilayers were imaged in both
contact and tapping mode using a MultiMode atomic force microscope
with NanoScope IIIa controller (Digital Instruments, Santa-Barbara,
CA).  Oxide-sharpened silicon nitride AFM probes (Digital Instruments,
Santa-Barbara, CA) with nominal spring constants of 0.06 N/m and 0.12
N/m were used.  To ensure that the force was kept minimal during
scanning, the force was frequently decreased until the tip left the
surface and was subsequently slightly increased until just regaining
contact.  The scan rate was 5-8 lines per second (Hz) for contact mode
images and 1-3 lines per second (Hz) for tapping mode images.  All
images have 512 $\times$ 512 pixels.  Analysis of AFM images was
performed with non-commercial software WSxM\copyright~(Nanotec
Electr{\'o}nica, http://www.nanotec.es).

\section{Results}
\subsection{Atomic Force Microscopy}
\begin{figure*}[htb!]
	\begin{center}
	\includegraphics[width=\textwidth]{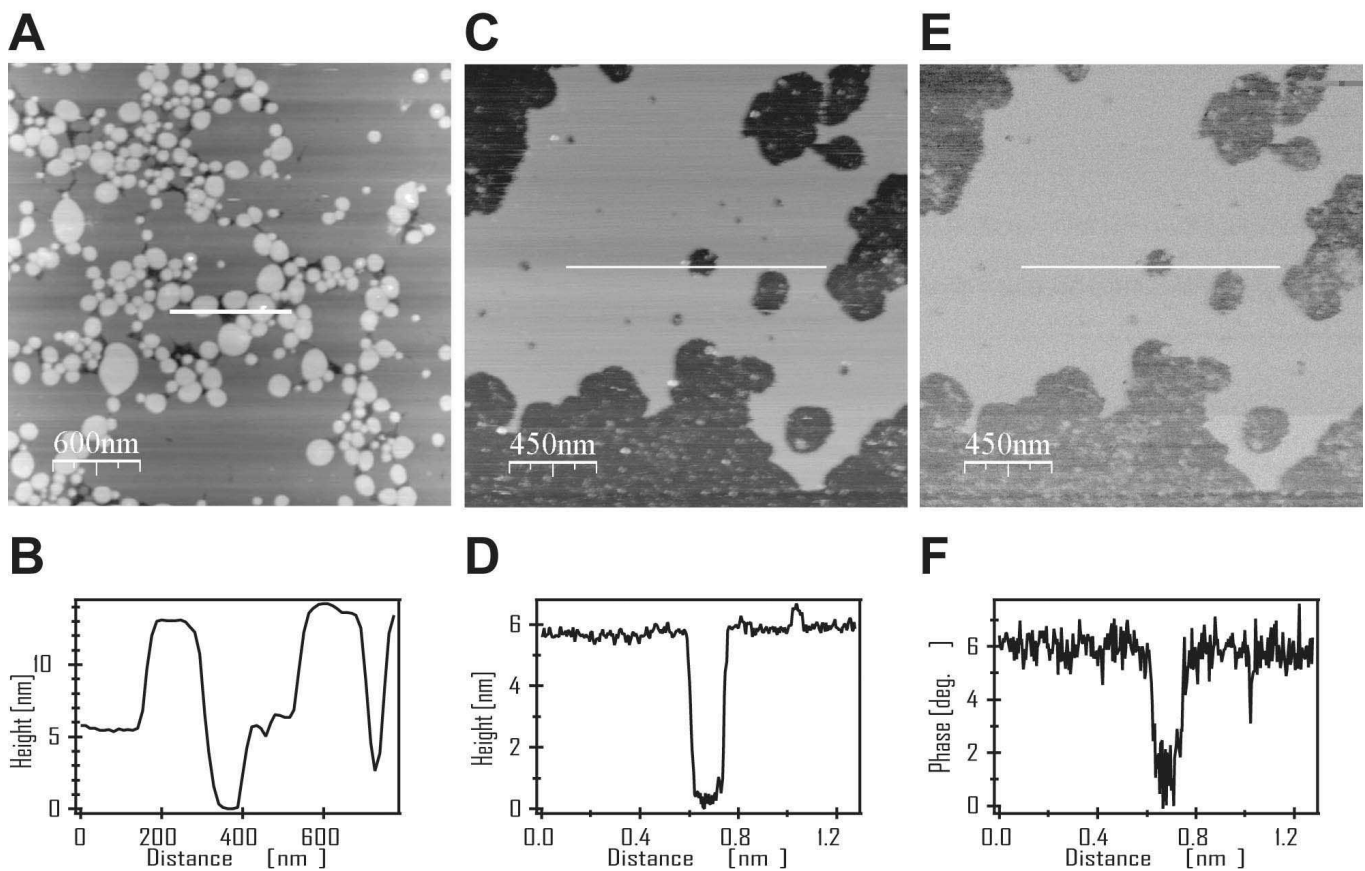}
	\caption{~DPPC and DLPC mica supported bilayers without
	peptides added.  (\textit{A}) $3\times3~\mu$m scan size height
	image of DPPC membrane obtained operating the AFM in contact
	mode and (\textit{B}) cross-section plot along the white line
	on the height image (\textit{A}), from which thicknesses of
	the first and the second bilayers are $\sim6$~nm and
	$\sim8$~nm.  Simultaneously recorded tapping mode height
	(\textit{C}) and phase (\textit{E}) images of a DLPC membrane
	with $2.2\times2.2~\mu$m scan sizes where white lines mark the
	position of cross-section profiles shown in (\textit{D}) and
	(\textit{F}), respectively, and corresponds to the same scan
	line and the same position on the sample surface.  All images
	have 512 scan lines and 512 points per line.}
	\label{Figure1}
	\end{center}
\end{figure*}

Without membrane active peptides, mica-supported lipid bilayers in
gel or fluid phases display a very flat surface when visualized with
AFM resolutions. It can show up rare structures caused by defects in
the crystal structure of the mica and by contamination or partial
fusion of lipid bilayers. A random example indicating different
modes of lipid membrane fusion on mica is shown in~\figref{Figure1}.
Part \textit{A} of this figure presents an AFM height image of a DPPC
membrane supported on mica. Very dark regions of the image indicate
the height level of the mica support. Hence they reveal parts of mica
which were not covered by lipid. Such defects may be due either to a
too short incubation time or to a too low lipid concentration in the
droplet of DPPC sample placed on the freshly cleaved mica surface.
Very bright areas show small pieces of a second lipid bilayer laying
on the top of the first one. This view is supported by the height
profile shown in  part \textit{B} of~\figref{Figure1}. The height of
the first DPPC bilayer in the gel phase is between 5 and 6 nm which
is in accordance with the accepted thickness of DPPC
membranes~\cite{Kaasgaard2003}. The distance between the upper
surfaces of the first and the second bilayers is by about 2 nm
larger, demonstrating that it involves the thickness of a water layer
between both membranes \cite{Nagle1980,Makowski1984}.

Parts \textit{C} and \textit{E} of \figref{Figure1} present a height
image and a phase image, respectively, of the same area of a mica
supported DLPC membrane. Contrary to DPPC, with main phase transition
temperature $T_m=41.6$$^{\circ}$C, the DLPC membrane ($T_m=-2.1$$^{\circ}$C) is
in the fluid phase at room temperature. The DLPC bilayer appears as a
flat leaflet in the height image. The height profile (\textit{D},
\figref{Figure1}) indicates a membrane thickness of somewhat less
than 6~nm, as determined from the height difference between the
bilayer surface and the mica support surface at a membrane defect.

In the phase image of the DLPC membrane area (\textit{E},
\figref{Figure1}), which has been recorded simultaneously with the
tapping mode height image, brighter coloring indicates larger phase
shifts. Hence the image shows larger phase shifts in the tapping
oscillations when probing the soft membrane instead of the stiff mica
support. Here phase imaging techniques will be used to highlight
lateral membrane structures, in particular to study the effect of
peptides upon the lateral distribution of viscoelastic membrane
properties.
\begin{figure*}[htb!]
	\begin{center}
	\includegraphics[width=\textwidth]{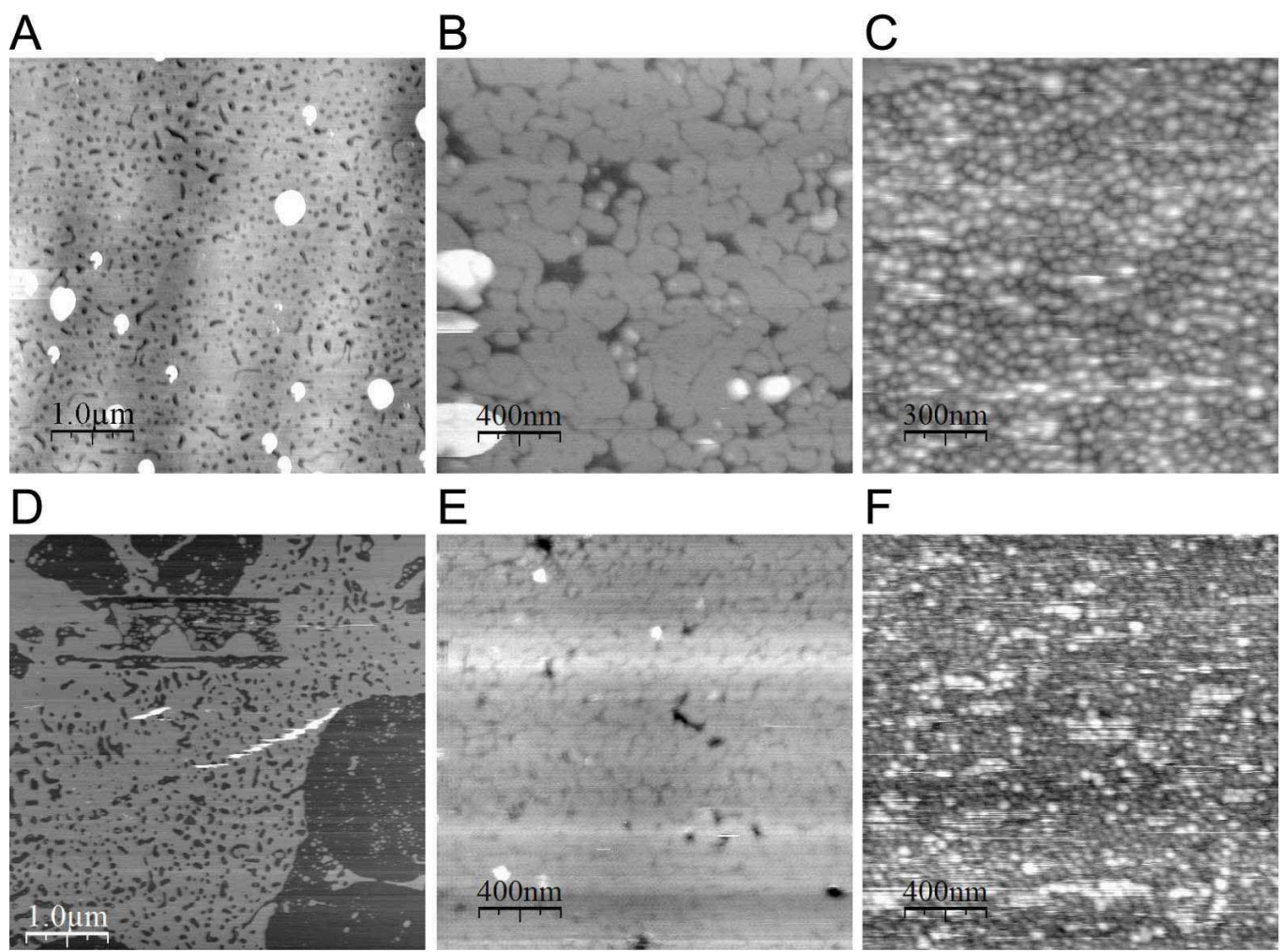}
	\caption{~Atomic force micrographs of lipid membranes
	deposited on mica in the presence of pore forming peptides.
	(\textit{A}) Height image (5$\times$5~$\mu$m scan size) of
	1~mol~\% alamethicin in DPPC membranes in the gel state at
	25--30$^{\circ}$C.  Dark spots on the brighter surface of the
	DPPC membrane represent the mica surface, which is accessible
	for AFM tip via alamethicin induced transmembrane defects.
	White regions in all images correspond to the second sheet of
	lipid membrane which lies on the top of the first one.  Height
	of DPPC membrane slightly varies across the image, which is
	caused by imperfectly adjusted scanning parameters.
	(\textit{B}) Height image (2$\times$2~$\mu$m scan size) of
	1~mol~\% melittin in DPPC membrane in the gel state.  Dark
	regions represent the mica surface, and the bright regions
	represent DPPC membrane segments.  Linelike depressions in the
	membrane can be seen.  (\textit{C}) Height image
	(1.5$\times$1.5~$\mu$m scan size) of 1~mol~\% melittin in DLPC
	membrane in the fluid state.  A highly developed net of
	depressions on the membrane surface is clearly demonstrated.
	A set of figures in the bottom panel (\textit{D}, \textit{E}
	and \textit{F}) represents images obtained under the same
	conditions and for the same lipid/peptide systems as shown
	above (for \textit{A}, \textit{B} and \textit{C},
	respectively) but from completely different sample
	preparation.}
	\label{Figure2}
	\end{center}
\end{figure*}
When 1~mol~\% of alamethicin is incorporated into a DPPC bilayer the
surface of the membrane appears evenly perturbed in the AFM height
images. As shown by parts \textit{A} and \textit{D}
of~\figref{Figure2} where results for two different sample
preparations are presented, predominantly circularly shaped
transmembrane defects result. Some elongated defects exhibit smooth
round kinks, as will be discussed with more details below (dark areas
in~\figref{Figure2} \textit{A}, \textit{D}). We assume these defects
to indicate alamethicin induced membrane holes, filled with water. We
were also able to directly observe melittin induced hole formation in
phosphocholine lipid bilayers. Samples of DPPC and DLPC were
investigated between 25$^{\circ}$C and 30$^{\circ}$C, both with and without
1~mol~\% of melittin added. Two different lipids were used, because
the AFM was not provided with a sufficiently stable temperature
control. The temperature of measurements was, therefore, slightly
above 25$^{\circ}$C for both lipids. DPPC at those temperatures is in gel
phase, whereas DLPC is in fluid phase. Some representative results
are again shown for two different samples in parts \textit{B},
\textit{E} as well as \textit{C}, \textit{F}, respectively,
of~\figref{Figure2}.

In contrast to the alamethicin, melittin forms in gel and fluid lipid
bilayers (DPPC and DLPC, respectively) a widespread net of
prolongated transmembrane defects with close spacing of defects
sides. Already at such small amount of peptide the bilayer looks
disintegrated. As another noticeable result, the pattern of melittin
induced holes in DPPC and DLPC bilayers is strikingly different. In
the fluid phase DLPC bilayer melittin develops a net of transmembrane
defects of high density and displays a more disordered structure than
in the gel phase DPPC bilayer, containing the same amount of peptide.

\begin{figure}[htb!]
	\begin{center}
	\includegraphics[width=8cm]{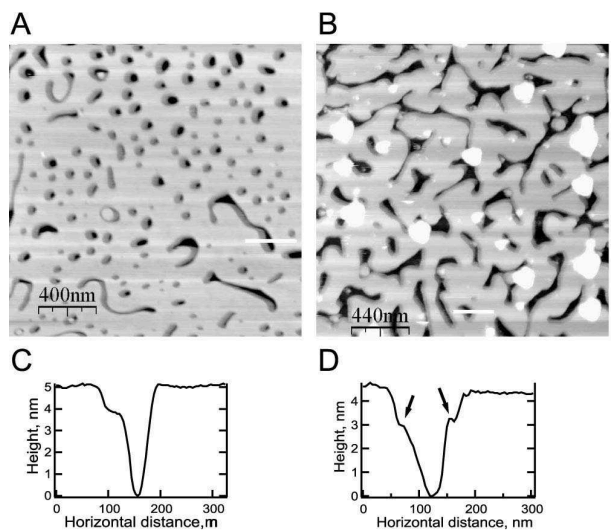}
	\caption{AFM images of gel DPPC membranes supported on mica in
	the presence of alamethicin of different concentrations.
	(\textit{A}) Height image (2$\times$2~$\mu$m scan size) of
	1~mol~\% alamethicin in a gel DPPC membrane.  Selected section
	of the same sample as in~\figref{Figure2}\textit{A}, but with
	smaller scan range.  Dark regions represents the mica surface
	at the bottom of water filled defects in the bright coloured
	DPPC membrane.  (\textit{B}) Height image
	(2.2$\times$2.2~$\mu$m scan size) of 4~mol~\% alamethicin in
	gel DPPC membrane.  Very bright regions correspond to the
	particles of great height which are most probably pieces of
	titanium left after sample sonification.  (\textit{C}) and
	(\textit{D}) are the cross-section height profiles along white
	marker lines extracted across selected alamethicin induced
	transmembrane defects in the figures \textit{A} and
	\textit{B}, respectively.  Black arrows point to stepwise
	changes in the thickness of the gel DPPC membrane close to the
	peptide hole.  Measured height difference correlates well with
	height difference between the thickness of a DPPC membrane in
	gel and fluid phase.}
	\label{Figure3}
	\end{center}
\end{figure}

When the samples of DPPC bilayers containing 1~mol~\% of alamethicin
shown in~\figref{Figure2}\textit{A} was scanned on a large scale, an
interesting defect structure emerged (\figref{Figure3}\textit{A}). In
this image the diversity of defect formation by alamethicin is
demonstrated in details. Most holes appear preferentially as round
transmembrane defects of different sizes. There is also a small
number of peptide induced elongated holes, some of which appear as
roundish kinks of specific radius.

Another interesting feature of alamethicin induced defects exists at
the membrane-hole interface.  Close to almost all peptide induced
holes there exist kinds of shells, exhibiting a lower height (darker
areas in the height images) than the undisturbed bilayer.  From the
height profile across one of the holes (\figref{Figure3}\textit{C}) we
found the height of the shell in the range of 3 to 4~nm, whereas the
bilayer thickness amounts to 5--6~nm.  This feature appears in all
alamethicin-containing samples independent of scan direction.  It does
not show up at the interface of pure lipid membranes as shown in Fig.
\figref{Figure1}.  We therefore concluded that these regions at the
pore interfaces are no experimental artifacts but a rather a result of
the interaction of the peptides with the lipid membrane in its
environment.  As mentioned before, the latter bilayer height is in
good agreement with literature values for the thickness of gel phase
DPPC lipid bilayers~\cite{Tamm1985,Mou1994,Heimburg1998}.  As the
melting temperature of DPPC bilayers without peptide added is around
41$^{\circ}$C the alamethicin containing membrane will be in the gel phase
at room temperature.  The height of the shells close to the
alamethicin induced holes corresponds to the thickness of fluid DPPC
membranes~\cite{Heimburg1998}.  Hence these are strong indications
that these shells around the peptide induced holes are nanoscopic
fluid phase lipid domains, spontaneously formed to reduce the
structural mismatch.  The length of alamethicin helices is about
3.5~nm which is significantly less than the thickness of DPPC
membranes in the gel state.  In order reduce at the peptide-lipid
membrane interface the unfavourite interactions between the DPPC
hydrophobic chains and water, phospholipid molecules near alamethicin
tend to exist in the fluid phase with reduced membrane thickness
rather than in the gel phase with stretched hydrophobic chains.

Increasing the alamethicin concentration in DPPC bilayers up to
4~mol~\%, the peptide do no longer form circularly shaped defects but
rather such with an elongated, branched, and irregular shape
(\figref{Figure3}\textit{B}). Quite remarkable, the height reduction
in the membrane close to alamethicin induced holes are also detected
at this fourfold higher peptide concentration, as clearly revealed by
the height profile (\figref{Figure3}\textit{D}). These findings
demonstrate the similarity in the behaviour of membranes with
different peptide concentration.
\begin{figure}[htb!]
	\begin{center}
	\includegraphics[width=8cm]{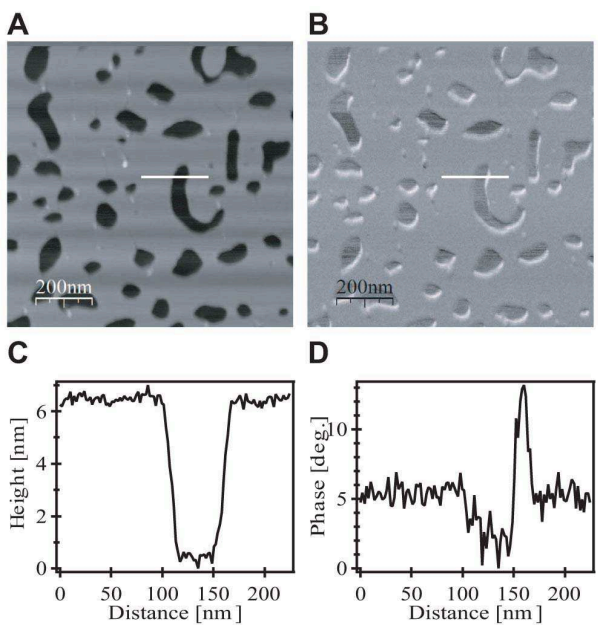}
	\caption{Atomic force microscopy scans of gel DPPC lipid
	membrane with 1~mol\% of alamethicin supported on mica.
	(\textit{A}) Height and (\textit{B}) phase images,
	respectively, from the 1~x~1~$\mu$m scan of the selected area
	from the~\figref{Figure2}\textit{D}.  The white line in both
	images depicts the cross-setion height (\textit{C}) and phase
	(\textit{D}) profiles.  Brighter areas close to peptide
	induced membrane defects depict greater phase shifts in AFM
	tip vibrations as can be seen also in phase cross-section plot
	(\textit{D}).  }
	\label{Figure4}
	\end{center}
\end{figure}
In~\figref{Figure4} height and phase contrast images of a DPPC
membrane containing 1~mol~\% alamethicin are presented.  Again both
images were simultaneously recorded in the tapping mode.  The phase
image clearly reveals some sites close to the holes with a bright
contrast.  This feature is also illustrated by the phase profile
in~\figref{Figure4}\textit{D}.  This phase delay in the AFM tip
vibrations indicates that the membrane surface close to the peptide
induced hole is softer than the undisturbed lipid membrane.  Such
softening of the gel phase DPPC bilayer close to the peptide
aggregates shows the noticeable effect of alamethicin on the local
compressibility of the membrane.  This is again an indication of the
DPPC molecules near alamethicin to exist in an altered physical state
close to the transition range rather than in the gel phase that one
expects at room temperature.  The peptide induced softening of the
lipid membrane close to the holes is in conformity with Monte Carlo
simulations studies~\cite{Ivanova2003} which revealed increased
fluctuations of the lipid phase near the peptide clusters.  Larger
fluctuations correspond to a higher compressibility and are expected 
in the phase transition regime. In the calorimetric measurements 
in Fig. \figref{Figure6} it is shown that the presence of the 
peptides lowers melting points (see below).  Although the
height image (\figref{Figure4}\textit{A}) as well as the height
profile (\figref{Figure4}\textit{C}) do not reveal obvious changes in the
membrane thickness near the peptide clusters, this feature is clearly 
present in other
alamethicin/DPPC samples (\figref{Figure3}). The slightly different 
appearence of identical sample preparations is likely due to small
temperature variations between the different experiments because the
atomic force microscopes were not provided with a temperature control.
Obviously, in the example shown in~\figref{Figure4} the lipid, as
judged from the membrane thickness, is still in the gel phase but
there exist already enhanced fluctuations between the two states as 
expected in the phase transition regime.  For
this reason there exist regions near the peptide clusters in which the
compressibility of the membrane is enhanced, leading to the more
detailed phase image.
\begin{figure}[htb!]
	\begin{center}
	\includegraphics[width=8cm]{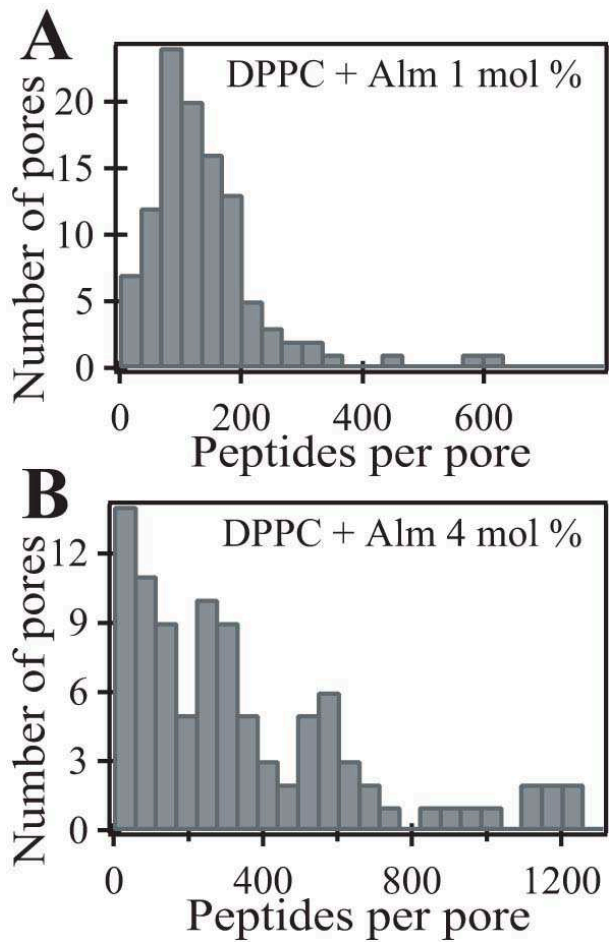}
	\caption{(\textit{A}) and (\textit{B}) are histograms of
	distribution of peptide aggregation number per pore calculated
	for DPPC membranes with 1 and 4~mol~\% of alamethicin from AFM
	height images~\figref{Figure3}\textit{A} and \textit{B},
	respectively.  Number of peptides per pore was obtained from
	dividing the measured pore perimeter by known from literature
	peptide-to-peptide distance of $\sim$11\r{A}.  The broadness
	of histogram bars equals about 33.}
	\label{Figure5}
	\end{center}
\end{figure}
In order to evaluate the hole sizes as induced by alamethicin in gel
phase DPPC bilayers one can argue in terms of an "aggregation
number", i.e., the number of peptides in the cluster around a hole
filled with water. Measuring the length of the inner perimeter of
every hole in the DPPC bilayers shown in \figref{Figure3}\textit{A}
and \ref{Figure3}\textit{B}, we recorded the number of pores
exhibiting a certain perimeter length. Assuming the
peptide-to-peptide distance to equal
$\sim$11~\r{A}~\cite{He1996,Spaar2004},  it is possible to estimate
the number of peptides that form a hole by dividing the perimeter of
the hole by the peptide-to-peptide distance mentioned above. In doing
so we obtained a distribution of aggregation numbers shown by the
histograms in~\figref{Figure5}\textit{A} and \ref{Figure5}\textit{B}
which are plotted for DPPC bilayer samples containing 1 and 4~mol~\%
of alamethicin, respectively. From the histograms we obtained a ratio
1:3 for the total numbers of aggregated alamethicin molecules in
membranes with 1~mol~\% and 4~mol~\% peptide concentrations,
respectively. This ratio is larger than the 1:4 concentration ratio,
probably, at least in parts, due to the experimental limitations.
Height reductions with an area smaller than $\sim$100~nm$^2$ where
not taken into account, since the used AFM tips had a radius of
curvature of about 10~nm. Peptide-induced holes, with dimensions
smaller than the characteristic size of the AFM tip, could indeed be
detected in height images, however, on such events  the AFM tip could
not reach the bottom of the hole, namely the mica surface.
Therefore, the hole perimeter can only be estimated from the height
images.

\subsection{Differential Scanning Calorimetry}
\begin{figure}[htb!]
	\begin{center}
	\includegraphics[width=8.0cm]{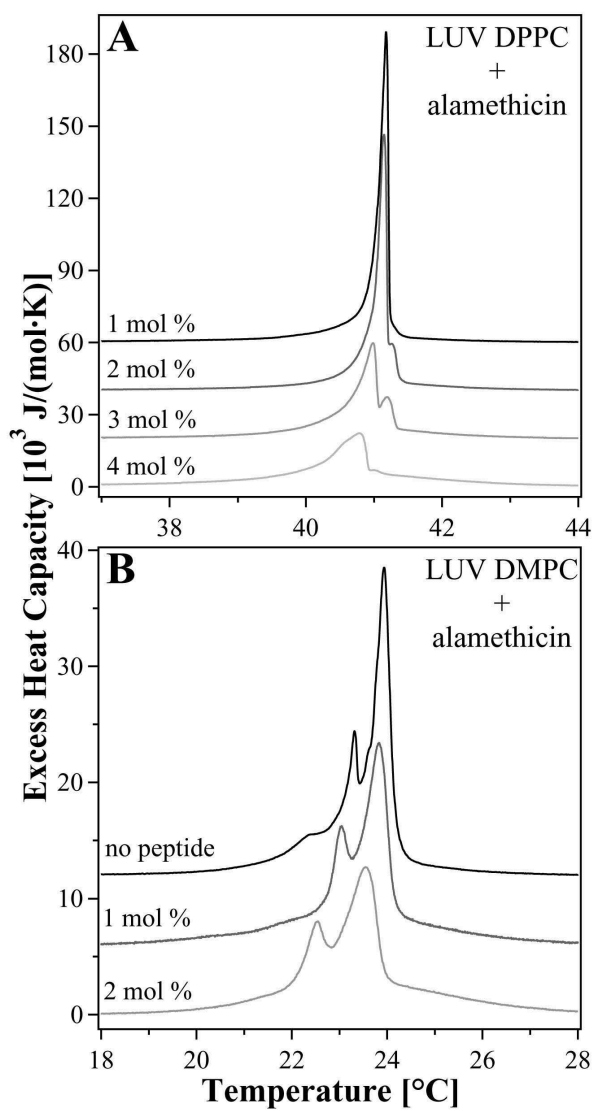}
	\caption{Alamethicin in phosphocholine lipid membranes.
	(\textit{A}) Heat capacity profiles of alamethicin in DPPC
	large unilamellar vesicles with 1, 2, 3 and 4~mol~\% of
	peptide.  (\textit{B}) Heat capacity profiles of alamethicin
	in DMPC large unilamellar vesicles with 0, 1 and 2~mol~\% of
	peptide.  In all DSC-scans addition of alamethicin results in
	a minor shift to lower temperatures and in a broadening of the
	main transition peak.  For a better data representation curves
	were shifted for a constant offset along the heat capacity
	axis.}
	\label{Figure6}
	\end{center}
\end{figure}

Excess heat capacity profiles (further on called "$c_P$-profiles") of
large unilamellar vesicle (LUV) suspensions from DPPC and DMPC with
different amounts of alamethicin added are displayed
in~\figref{Figure6}, \textit{A} and \textit{B}. For a clear data
representation, plots of different alamethicin content were shifted
along the heat capacity axis using a constant offset. 
The heat capacity curve of the pure DPPC vesicles normally displays a
maximum at $\sim$41.6$^{\circ}$C, which is often called the main phase
transition temperature, $T_m$. In the $c_P$-profiles shown
in~\figref{Figure6}\textit{A}, $T_m$ and the shape of the
transition peak are increasingly affected with the concentration of
alamethicin
within the DPPC membranes. This finding is also demonstrated by the
$T_m$ values listed in~Tab.~\ref{tab:dsc_dmpc_dppc_alm}. The presence
of peptide is reflected in those heat capacity profiles by a slight
shift to lower temperatures accompanied by a small asymmetry at the
low temperature wing and a transition peak broadening. However, at
the highest measured peptide concentration of 4~mol~\%, the
transition temperature is decreased by less than 1K only. The lipid
melting enthalpy, $\Delta H$, which is determined as the area under
the
heat capacity-versus-temperature profiles, decreases slightly when
the alamethicin concentration in the DPPC vesicles increases up to
3~mol~\%. A similarly small influence of alamethicin on the
$c_P$-profiles of DMPC vesicles was also observed
(\figref{Figure6}\textit{B}). With DMPC membranes, however, $\Delta
H$ increases slightly with alamethicin content. A change in the
transition enthalpy of the DMPC system may, however, be caused by
increasing uncertainty to accurately
determine the base line.

In $c_P$-profiles of DPPC suspensions with alamethicin added one can
also observe that, at 2~mol~\% of peptide, a small second peak
appears at the high temperature slope of the main transition peak. It
is more developed in the $c_P$-profile for 3~mol~\% of peptide but
almost disappears with further increasing alamethicin content. The
heat capacity profile of pure DPPC LUVs normally does not show up a
splitting of the transition peak. However, in the case of extruded
DMPC vesicles this feature
exists~(see~\figref{Figure6}\textit{B}) and it is not noticeably
affected by the presence of alamethicin. The splitting of the peak in
$c_P$-profiles of pure DMPC LUV suspensions is believed to be related
to changes of the vesicle geometry in the lipid melting regime, by
analogy to a transition between lipid vesicles and a bilayer network
during lipid phase transition of DMPG dispersions, as detected in
electron microscopy experiments~\cite{Schneider1999}. Probably, such
appearance of an additional peak in $c_P$-profiles of alamethicin
containing DPPC membranes reflects morphological changes of free
lipid vesicles in solution, similar to the behaviour of DMPG
vesicles~\cite{Schneider1999}. This effect requires further
investigations.
\begin{figure}[htb!]
	\begin{center}
	\includegraphics[width=8.0cm]{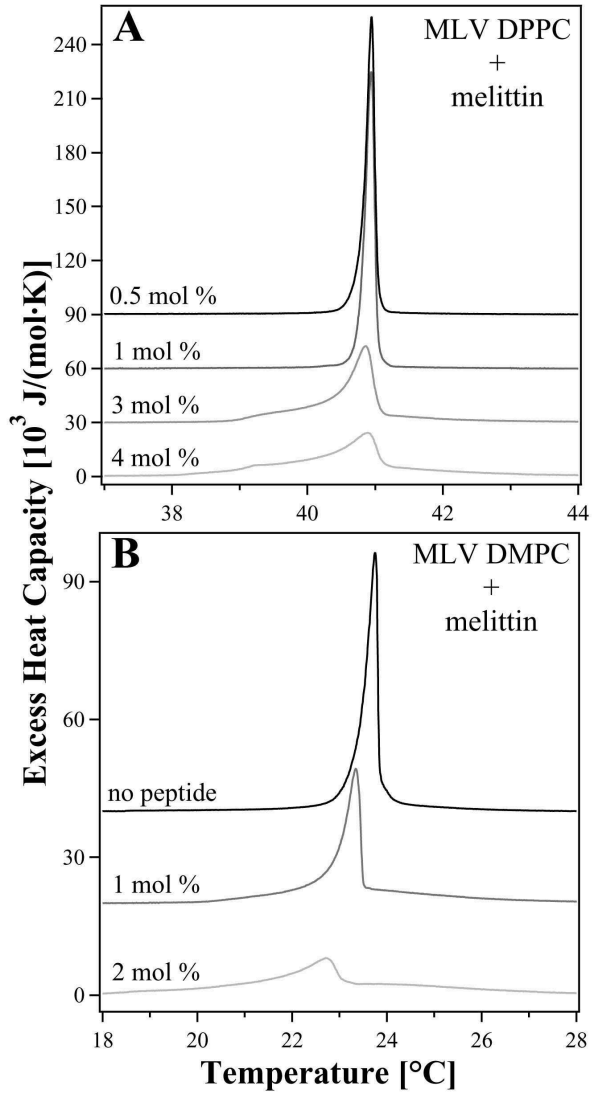}
	\caption{Melittin in phosphocholine lipid membranes.
	(\textit{A}) Heat capacity profiles of melittin containing
	DPPC multilamellar vesicles with 0.5, 1, 3 and 4~mol~\% of
	peptide.  (\textit{B}) Heat capacity profiles of melittin in
	DMPC multilamellar vesicles with 0, 1 and 2~mol~\% of peptide.
	When melittin is added the phase transition temperature tends
	to decrease accentuating the low-temperature shoulder which
	becomes more pronounced when the melittin concentration is
	increased.  For a better data representation curves were
	shifted for a constant offset along the heat capacity axis.}
	\label{Figure7}
	\end{center}
\end{figure}
Heat capacity profiles for melittin containing multilamellar vesicles
from DPPC and DMPC are displayed in~\figref{Figure7}. These vesicles
do not show the splitting as observes with the LUVs. When melittin is
added the phase transition temperature tends to decrease
(Tab.~\ref{tab:dsc_dmpc_dppc_alm}). Again, however, the reduction of
$T_m$ with peptide content is small. More evident is the
low-temperature shoulder in the $c_P$-profiles which becomes more
pronounced when the melittin concentration is
increased~\figref{Figure7}. Similar to alamethicin, melittin causes
only small variations in the enthalpy changes $\Delta H$ at the chain
melting temperature (Tab.~\ref{tab:dsc_dmpc_dppc_alm}).\\

\begin{table}
    \begin{center}
	\begin{tabular}{lcccc}
	    \hline\\
	    Peptide&\multicolumn{2}{c}{DPPC}&\multicolumn{2}{c}{DMPC}\\
	    content& $T_m$ & $\Delta H$ & $T_m$ & $\Delta H$\\
	    \hline\\
	    0.0 mol\qquad \% & 41.6[*] & 39.3 [*] & 23.6 & 22.2\\
	    1.0 mol\qquad \% & 41.18 & 34.8 & 23.4 & 22.5\\
	    2.0 mol\qquad \% & 41.17 & 34.6 & 22.9 & 23.1\\
	    3.0 mol\qquad \% & 40.98 & 29.7 & --- & ---\\
	    4.0 mol\qquad \% & 40.79 & 32.4 & --- & ---\\
	    \hline
	    \label{tab:dsc_dmpc_dppc_alm}
	\end{tabular}
	\caption{Melting temperatures $T_m$~($^{\circ}$C) and enthalpy
	changes $\Delta H$~(kJ/mol) determined from $c_P$-profiles of
	alamethicin containing membranes shown in~\figref{Figure6}. ([*] 
	values taken from \cite{Heimburg1998})}
    \end{center}
\end{table}

\section{Discussion and Conclusions}
Atomic force microscopy offers a valuable tool to visualize the
spatial organization of peptide defects in membranes.  For alamethicin
containing bilayers we found that the peptide induces holes in gel
state DPPC membranes, the hole structure depending on the peptide
concentration.  Since in the literature \cite{Fox1982,Gennis1989}
pores are usually described as being very small we losely refer to
these holes as `defects'.  However, we have no evidence that the
principle features of the peptide pores and our defects are different.
An increase of alamethicin content up to 4~mol~\% in DPPC bilayers
leads to the formation of bigger and more irregularly shaped
transmembrane defects in comparison to the four times lower peptide
concentration, at which alamethicin aggregates preferentially into
smaller and almost circularly shaped holes.  Such pore formation
dependence upon the peptide content in membranes fits well to the
commonly accepted barrel-stave model of alamethicin pores, in which
multilevel conductance of alamethicin channels is assigned to a
varying number of peptide molecules participating in the pore
formation~\cite{Hall1984,Duclohier1992,Laver1994,Cantor2002,Duclohier2001}.
From calculations of pore perimeters in AFM height images
(see~\figref{Figure5}, \textit{A} and \textit{B}) we found that, at
higher concentrations, alamethicin forms holes with a greater number
of aggregated peptide molecules per pore.  For 1~mol~\% of alamethicin
the peptide aggregation number varies from 30 to 300 alamethicin
molecules per hole with a maximum in the size distribution around 80,
while at 4~mol~\% of alamethicin in DPPC membranes the number of
peptide molecules forming a single aqueous defect can reach a value of
up to 1200.  However, such large numbers of alamethicin molecules,
aggregated into a transmembrane hole, is not characteristic for
alamethicin.  For example, it was shown previously by
neutron-scattering experiments~\cite{He1996} that in DLPC membranes
pores are made of 8--9 monomers, with a water pore diameter
of~$\sim$18~\r{A} and with an effective outside diameter
of~$\sim$40~\r{A}.  In diphytanoyl phosphatidylcholine membranes, the
pores are made of \textit{n}~$\sim$11 peptide molecules, with a water
pore $\sim$26~\r{A} in diameter and with an effective outside diameter
of $\sim$50~\r{A}~\cite{He1996}.  On the other hand, it was suggested
that a barrel-stave model with a water pore greater than 3.0~nm in
diameter would consist of a bundle of 12 or more peptide helices,
which is most likely unstable against shape
deformation~\cite{Yang2001}.  The observation of a wide distribution
of the alamethicin aggregation number in our AFM experiments may
suggest an effect of the mica surface on which the membranes were
spread during the experiment.  Contact with the mica crystal surface
may stabilize alamethicin pores and holes.  It is known that the
direct contact of a membrane with mica can alter the occurance of
ripple structures in pure lipid membranes \cite{Leidy2002}.  Therefore
an influence of the mica on the present results cannot completely
beruled out.  Additional experiments of multilayered membranes, in
which the influence of the supporting surface is reduced, may be
performed in future, in order to study this phenomenon in more
details.  However, the formation of big pores by alamethicin molecules
may be the native property of this peptide related to its antibiotic
action.  Alamethicin aggregates, consisting of a large number of
peptide molecules, are difficult to detect utilizing
neutron-scattering, while atomic force microscopy provides a
possibility to inspect those peptide pores.  However, pores with a
number of peptide molecules smaller than about 30 (diameter smaller
than 10~nm) can not be resolved with AFM techniques because of the
finite radius of curvature of the tip.

In melittin containing membranes at 1~mol~\% peptide concentration
transmembrane pores are formed in both DPPC and DLPC mica supported
bilayers.  Since DPPC membranes are in the gel phase at room
temperature and DLPC membranes in the fluid phase, we conclude that
the capability of melittin to form transmembrane defects does not
depend on the state of the mica supported bilayer.  The shape of the
defects in both phases, however, is different.  In DPPC bilayers
melittin molecules form elongated line-like structures, likely
reflecting the high degree of ordering of the lipid matrix in the gel
phase (similar structures were found for gramicidin A in the gel
phase, \cite{Ivanova2003}).  In DLPC membranes melittin molecules
aggregate to a highly disordered branched net of pores, in conformity
with the less ordered fluid phase of the lipid membrane.

Another important result of this study of peptide containing lipid
membranes is the existence of nanoscopic domains of lower height in
close vicinity to the peptide induced pores. In AFM images of the gel
DPPC bilayers with alamethicin and also with melittin we found local
height depressions 
close to the peptide pores. The height in these areas correspond to
the thickness of DPPC membranes in the fluid
phase~\cite{Heimburg1998}. This led us to the conclusion that, in a
gel lipid membrane, the peptide induces melting of the surrounding
lipids. Because of this melting a hydrophobic mismatch around peptide
molecules is reduced and hereby the free energy of the system is
minimized. From an free energy point of view alamethicin helices with
lengths of $\sim$3.5~nm and a predominantly hydrophobic
surface~\cite{Fox1982} in water, tend to be rather surrounded by a
DPPC membrane in the fluid phase with its thickness of $\sim$3.9~nm
than by a DPPC membrane in the gel phase with its larger thickness of
$\sim$4.8~nm~\cite{Heimburg1998}. Such an influence of peptide
aggregates on the thermodynamic state of contacting lipids was also
demonstrated in Monte Carlo (MC) simulations of peptide containing
membranes~\cite{Ivanova2003}. It has been shown by those studies that
the fluctuations of the lipid state at fixed temperature are higher
close to the peptide aggregates embedded into the gel lipid matrix,
which means a higher probability to find lipid molecules in a
fluid state. Melting of peptide coupled lipids occurs at lower
temperatures than of phospholipid membrane without peptides added.
This is reflected by the heat capacity profiles. $c_P$-profiles of
alamethicin containing membranes, shown in this work, demonstrate a
shift of the transition peak to lower temperatures as compared to
pure membranes. With increase of peptide content the shift of
$c_P$-profiles is also larger. Such trends in the measured heat
capacity curves are linked to the melting of certain fractions of
lipids at lower temperatures and to the formation of nanoscopic fluid
lipid domains close to the peptide-induced defects which were
detected in the AFM experiments. The presence of peptide molecules in
the membrane, which have a hydrophobic length shorter than chain
length of the surrounding lipids in the gel state tend to reduce the
energetic barrier for changing their state from gel to fluid.

In MC-simulations of gramicidin A containing membranes it has been
previously demonstrated~\cite{Ivanova2003} that peptide induced
shifts of $c_P$-profiles can be explained in terms of peptide
aggregation in lipid membranes. It was shown that considerable shifts
of the transition peak to lower or higher temperatures correspond to
a preferential aggregation of the peptide either in the gel or the
fluid phase, while the unchanged main transition temperature is
referred to the equally well aggregation of peptide in both phases.
For melittin containing membranes we found that the peptide induces a
shift of the transition peak in heat capacity profiles similar to
alamethicin. Applying the above mentioned analysis one can predict
melittin clustering (associated with pore formation) in both gel and
fluid lipid membranes. This is in agreement with the
AFM images presented in this work for DPPC (gel) and DLPC (fluid)
membranes containing 1~mol~\% of melittin. Melittin aggregation into
the transmembrane pores was observed in both membranes. We
demonstrated also that the structure of melittin pores depends on the
thermodynamic state of the membrane. In the case of fluid DLPC
bilayer, melittin develops a network of transmembrane pores of higher
density and more disordered structure as compared to gel DPPC
bilayers containing the same amount of peptide. Therefore the lytic
power of melittin should depend on the thermodynamic state of the
membrane.

Hence, the peptides in model and biological membranes can strongly
affect the local state of the system, and the effect of the peptide
may also depend on the overall state of the membrane. Peptides like
alamethicin tend to  increase the permeability of biological
membranes not only by forming water pores, but also by shifting the
surrounding bilayer to more disordered states which are more
permeable for small molecules. In turn, the state of the membrane can
be a regulating factor of the action of antibiotic peptides like
melittin,
influencing their capability to form transmembrane pores.

\section{Acknowledgements}
We are indebted to Dr.~T.~Sch\"affer, University of M\"unster, for
introducing one of us (V.O.) into atomic force microscopy and to
Prof.~T.~Bj\o rnholm for generously making the equipment of the
Nano-Science Center at the University of Copenhagen available for
this study. Thanks are also due to Dr.~M.~Konrad, MPI for Biophysical
Chemistry in G\"ottingen, for the synthesis of peptides. Financial
support from the graduate school "Neuronal Signalling and Cellular
Biophysics", Georg-August-University of G\"ottingen, is also
gratefully acknowledged.

\footnotesize
\renewcommand{\baselinestretch}{0.2}
\bibliographystyle{elsart-num}

\end{document}